%% file: Template.tex
\documentclass{article}
\usepackage{spconf,amsmath,graphicx}
\usepackage{xcolor} 
\usepackage{tabularray} 
\SetTblrInner{rowsep=0pt} 
\usepackage{hyperref}
\usepackage{comment}
\usepackage{balance} 
\usepackage{fancyhdr} 
\usepackage{setspace} 

\title{Speaker anonymization using neural audio codec language models}
\name{Michele Panariello,$^1$ Francesco Nespoli,$^2$ Massimiliano Todisco,$^1$ Nicholas Evans$^1$}
\address{$^1$EURECOM, France~~~~~~~~$^2$Microsoft UK}

\begin{document}
\pagestyle{fancy}
\fancyhf{}
\renewcommand{\headrulewidth}{0pt}
\fancyfoot[L]{{\color{gray} \footnotesize \textbf{©2023 IEEE.} Personal use of this material is permitted. Permission from IEEE must be obtained for all other uses, in any current or future media, including reprinting/republishing this material for advertising or promotional purposes, creating new collective works, for resale or redistribution to servers or lists, or reuse of any copyrighted component of this work in other works.}}
\input{symbols}
\maketitle
\begin{abstract}
The vast majority of approaches to speaker anonymization involve the extraction of fundamental frequency estimates, linguistic features and a speaker embedding which is perturbed to obfuscate the speaker identity before an anonymized speech waveform is resynthesized using a vocoder.
Recent work has shown that x-vector transformations are difficult to control consistently: other sources of speaker information contained within fundamental frequency and linguistic features are re-entangled upon vocoding, meaning that anonymized speech signals still contain speaker information.
We propose an approach based upon neural audio codecs (NACs), which are known to generate high-quality synthetic speech when combined with language models. NACs use quantized codes, which are known to effectively bottleneck speaker-related information: we demonstrate the potential of speaker anonymization systems based on NAC language modeling by applying the evaluation framework of the Voice Privacy Challenge 2022.
\end{abstract}
\section{Introduction}
\label{sec:intro}
\emph{Speaker anonymization} involves the task of processing a speech signal to conceal the identity of the speaker while retaining the spoken content and other para-linguistic attributes such as intonation and prosody.
As defined by the Voice Privacy Challenge~\cite{introducing_vp}, a speaker anonymization system should provide a certain trade-off between \emph{privacy protection} and \emph{utility preservation}. 
The former is measured by the difficulty of an attacker to recover the identity of the original speaker from an anonymized signal via automatic speaker verification (ASV).  The latter is assessed primarily by the reliability of an automatic speech recognition (ASR) system to transcribe the anonymized waveform, among other, secondary metrics such as pitch correlation and gain of voice distinctiveness~\cite{vpc2022}.

Most 
speaker anonymization systems are based on often-incremental changes to original work in~\cite{fang19}, which operates upon three distinct components 
extracted from an input waveform: an F0 curve encoding prosodic information; some form of linguistic features encoding the spoken content; an x-vector~\cite{ecapa} encoding the speaker identity.
The x-vector is perturbed to conceal the identity of the speaker and then fed to a vocoder with the other two components in order to synthesize an anonymized waveform.
This approach assumes that speaker information is contained entirely within the \mbox{x-vector}, even if this is known not to be the case~\cite{pierre_disent, f0_pierre}.
Residual speaker information captured in linguistic features and the F0 curve is re-entangled with the anonymized x-vector upon vocoding.  An x-vector extracted afresh from the anonymized utterance then still contains speaker information which can be used by an adversary to reverse the anonymization and re-identify the speaker~\cite{pierre_disent, dp}.
Other researchers~\cite{pierre_disent} have found that the level of speaker information contained in linguistic features can be reduced through their quantization. 

Motivated by their successful application to numerous audio synthesis tasks~\cite{audiolm,vall-e}, we have sought to exploit 
the potential of neural audio codec (NAC) language modeling
to design a speaker anonymization system that
better suppresses speaker information and hence provides an improved trade-off between anonymization and utility.
Such an approach is appealing since 
linguistic features are not used directly for waveform synthesis, as with previous approaches: instead,
they are used to infer a set of NAC acoustic tokens with a language model. These features are quantized and therefore have the potential to bottleneck speaker information and improve anonymization. The final waveform is synthesized by decoding the acoustic tokens with a decoder neural network.
We hope that the representational power of NACs should help to preserve speech quality and utility.

\section{Related work}
\label{sec:related_work}
In the following we describe some related research which provided motivation for the work presented in this paper.

\textbf{X-vector--based speaker anonymization -} The original x-vector--based pipeline introduced in~\cite{fang19} is the basis of much of the related work reported recently.
An approach to dispense with the intermediate acoustic model was proposed in~\cite{vpc2022}. 
More refined x-vector anonymization functions were proposed in later work~\cite{t11,T04}, with some~\cite{ohnn} achieving notable improvements to privacy protection levels, albeit under the assumption that the attacker does not have full knowledge of the anonymization system.
Whatever the approach, x-vector perturbation does not prevent speaker-related information contained in the F0 curve and linguistic features from 
\emph{leaking} into the anonymized waveform upon vocoding.
As a result, the x-vector which can be re-extracted from the anonymized waveform by a privacy adversary who wishes to re-identify the speaker tends to \emph{drift}
away from that at the vocoder input~\cite{drift2}.  
While the drift can be beneficial to anonymization,
it hinders the design of more effective anonymization functions and can also be inverted by an adversary to undo the privacy protection~\cite{drift1}.
Attempts to sanitize speaker information from linguistic features have also been explored, e.g.~\cite{dp} based on the concept of differential privacy, which reports improvements to privacy at the cost of reduced utility and pitch correlation.
The same issue was tackled in~\cite{pierre_disent} by means of feature quantization, which is shown to be effective as a speaker information bottleneck, though with a degradation to utility.
In this paper, we follow a similar approach, but propose a completely new synthesis pipeline in which we avoid the leakage of information between different speech components by design.

\textbf{NAC language modeling -} NACs were proposed recently for  audio compression~\cite{soundstream, encodec}. 
They consist in convolutional autoencoders that compress audio to low-bitrate, tokenized representations which support  high-fidelity decoding.
Due to their discretized nature, encoded representations can be modeled using techniques normally used for language-related tasks, such as transformers.
This idea was first introduced in~\cite{audiolm} for a variety of audio generation tasks, and tailored to text-to-speech (TTS) in~\cite{vall-e}.
A transformer is used to convert the input (be it audio or text) to a set of high-level \emph{semantic tokens}. 
These are then fed to another transformer which converts them into \emph{NAC acoustic tokens} which can be decoded to re-synthesize an audio signal.

The same technique can also be applied to voice conversion~\cite{audiolm,wang2023lmvc}, and is hence ideally suited to speaker anonymization.
NAC language models operate on quantized codes, which are known to be beneficial to privacy protection~\cite{pierre_disent}.
Moreover, such models appear to naturally disentangle linguistic information into semantic tokens, while encoding speaker information and recording conditions mostly in acoustic tokens~\cite{audiolm}.
Hence, we propose a speaker anonymization system whereby an input utterance is re-synthesized by means of a NAC language model. Semantic tokens are kept unchanged,
while acoustic tokens are substituted with those of a different speaker: the goal
is to preserve the linguistic content of the signal while suppressing information related to the original speaker.

\begin{figure}
    \centering
    \includegraphics[width=0.95\linewidth]{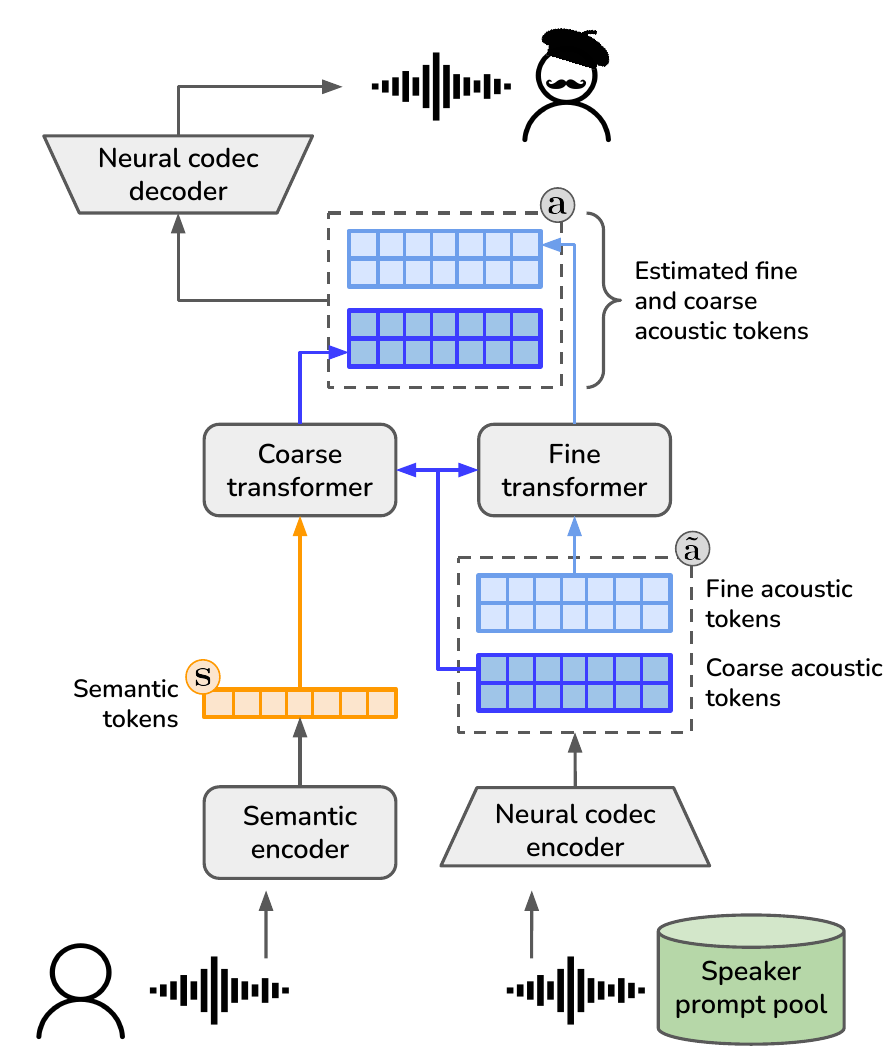}
    \caption{Diagram of the proposed anonymization system.}
    \label{fig:system}
\end{figure}
\pagestyle{empty} 

\section{Proposed approach}
\label{sec:model}

\subsection{Neural audio codec language modeling}
\label{sec:nac}
A diagram of the proposed system is shown in~\figref{fig:system}. Following~\cite{vall-e}, it is comprised of a semantic encoder, a NAC (encoder and decoder), a pair of transformers and a pool of speaker prompts.  They are described in the following.

\textbf{The semantic encoder} produces
high-level semantic representations of the input signal using a codebook of $N_S$ quantized embeddings.
The output is a sequence of integers $\s \in \{1, \dots, N_S\}^{T_S}$, where $T_S$ is the number of frames and where each  integer is a codeword index.

\textbf{The NAC} is an encoder-decoder architecture.
The encoder maps input waveforms to a quantized, compressed representation from which the decoder reconstructs a high-fidelity waveform.
Efficient compression is achieved with
a set of 
$Q$ hierarchical codebooks.
Lower level codebooks capture coarser waveform characteristics, while
finer details are captured by higher level codebooks.
Following~\cite{audiolm, vall-e}, we refer to the first $Q_C$ codebooks as `coarse codebooks', and to the last $Q - Q_C$ codebooks as `fine codebooks'.
All have $N_Q$ codewords so that the output of the encoder is $\prompt \in \{1, \dots, N_Q\}^{Q \times T_A}$, where $T_A$ is the number of frames into which the input is divided.

\textbf{The coarse and fine transformers}
estimate a set of acoustic tokens $\prediction$ from a prompt of input semantic tokens $\s$ and
acoustic tokens $\prompt$. 
Essentially, the transformers attempt to predict what   semantic information should `sound like' in the domain of  quantized acoustic tokens.
The coarse transformer autoregressively predicts coarse acoustic tokens, i.e.\ the codewords belonging to the coarse codebooks. More specifically, for frame $t$, the transformer predicts the probability distribution of token $\prediction_{q,t}$
conditioned on the following elements: the semantic prompt $\s$, the coarse tokens from the acoustic prompt $\prompt_{<Q_C, :}$, and all previous predictions.\footnote{In practice, the sequence upon which to perform regression is flattened to $(\s, \prompt, \prediction_{1,1}, \prediction_{2,1}, \dots, \prediction_{Q_C, 1}, \prediction_{1,2}, \prediction_{2,2}, \dots, \prediction_{Q_C,2},\dots).$ See~\cite{audiolm, vall-e} for more details.}
The modeled distribution is therefore
\begin{equation}
	p\left(\prediction_{q, t} \middle| \s, \prompt_{<Q_C, :}, \prediction_{<Q_C, <t}, \prediction_{<q, t} \right) 
\end{equation}
for $q \in [1, Q_C]$.
The fine transformer is instead non-autoregressive. It estimates the tokens of codebook $q$ using all tokens belonging to codebooks $<q$ and all tokens of the acoustic prompt $\prompt$, thus modeling the distribution
\begin{equation}
    p\left(\prediction_{q, :} \middle| \prompt, \prediction_{<q, :} \right)
\end{equation}
for every $q \in [Q_C+1, Q].$
Once the acoustic tokens $\prediction$ have been predicted for all codebooks $q \in [1,Q]$, they can be input into the decoder to synthesize an anonymized waveform.

\textbf{The pool of speaker prompts} is a set of acoustic tokens extracted by the NAC encoder from utterances belonging to a set of external speakers. Those speakers are referred to as \emph{pseudo-speakers}, since they replace the original speaker in the anonymized utterance.
As suggested in~\cite{audiolm}, acoustic tokens, especially the coarse tokens, can capture information related to the speaker identity.
We use them to perform voice conversion, as detailed in the following.

\subsection{Anonymization technique}
\label{sec:anon_technique}
A set of semantic tokens $\s$ is first extracted
from the input utterance.
These tokens encode the high-level spoken content.
Their quantization helps to suppress speaker-related information.

A pseudo-speaker is chosen by randomly selecting an acoustic prompt $\prompt$ from the speaker prompt pool. 
Anonymization can be performed at either speaker or utterance levels.
At the
\emph{speaker level}, anonymization is performed using the same speaker prompt for each utterance corresponding to
one specific speaker.
In contrast, for \emph{utterance level} anonymization, a speaker prompt is selected at random for each utterance.
While several anonymization systems include techniques to synthesize fictitious voices~\cite{pierre_disent,t11,T04,ohnn}, here we use real voices as pseudo-speakers to focus our analysis on the intrinsic anonymization capability of the NAC language model.

Prompted with $\s$ and $\prompt$, the coarse and fine transformers generate a set of acoustic tokens $\prediction$ which reflect the semantic information of the original utterance, but the acoustic characteristics of the pseudo-speaker.
Acoustic tokens $\prediction$ are fed to the
decoder which synthesizes the anonymized output waveform.

\input{table_res}

\section{Experimental setup}
\label{sec:experimental_setup}
Our codebase is branched from Bark,\footnote{The original source code is available at \url{www.github.com/suno-ai/bark}, though we built our system from the port included in the CoquiTTS library available at \url{www.github.com/coqui-ai/TTS}. Our source code, as well as audio samples, are available at \url{www.github.com/eurecom-asp/spk_anon_nac_lm}.} an open-source, \mbox{NAC-based} TTS system which is very similar to VALL-E~\cite{vall-e}.  The modules described in \secref{sec:nac} are all taken from Bark.
The semantic encoder has a HuBERT backbone~\cite{hubert} and a LSTM~\cite{lstm_oh_my_god_am_i_really_citing_this} back-end which predicts the semantic token associated to the HuBERT feature vector output at each frame. The semantic dictionary is of size $N_S = 10000$.
The coarse and fine transformers are 12-layer GPT-like models~\cite{gpt2}, each having $Q=8$ different codebooks with $N_Q = 1024$ codewords. The first $Q_C = 2$ codebooks are considered coarse.
The NAC is EnCodec~\cite{encodec}.
The difference between Bark and our system is that, being a TTS model, Bark estimates semantic tokens $\s$ corresponding to an input text using a further \emph{semantic transformer}. In our implementation, we bypass the semantic transformer and use ground truth semantic tokens from the input waveform thereby providing for voice conversion instead of TTS. With this setup, we are able to use pre-trained Bark modules off-the-shelf, without the need for any training.

We adopt the Voice Privacy Challenge 2022 protocol~\cite{vpc2022} for evaluation. 
The test set comprises subsets of the \emph{LibriSpeech}~\cite{librispeech} and \emph{VCTK}~\cite{vctk} databases. 
The pool of speaker prompts is taken from the Bark voice library. 
It consists of 130 utterances collected from speakers of different gender and nationality.\footnote{\url{www.github.com/suno-ai/bark/tree/main/bark/assets/prompts/v2}}
The threat model is the \emph{semi-informed} attack described in~\cite{vpc2022}. 
Trial utterances are anonymized at the \emph{speaker level}.
The attacker is assumed to have access to the anonymization system.
They anonymize a set of external data (librispeech-clean-360) at the \emph{utterance level} and use it to train an ASV system (a TDNN with a PLDA back-end~\cite{tdnn}).
They also have access to original (non-protected) enrollment utterances which they anonymize at the \emph{speaker level}.
The attacker thus has enrollment and trial utterances both of which are anonymized and uses an ASV system 
to verify whether they correspond to the same speaker. 
The attacker has no knowledge of which pseudo-speaker was used for anonymization on the test utterance and will hence likely select a different pseudo-speaker to anonymize the enrollment utterance.
The privacy metric is the resulting EER estimated from a large number of ASV trials; a higher EER corresponds to better privacy. 
Utility is assessed by training an ASR system on the same anonymized version of librispeech-clean-360 and by estimating the word error rate (WER) from anonymized test data; a lower WER corresponds to better utility. 
Two additional metrics are defined in the VoicePrivacy Challenge evaluation plan~\cite{vpc2022}.
The first is the F0 curve correlation $\boldsymbol{\rho}^{F0}$ between original and anonymized utterance which is used as a measure of prosody preservation.
The second is the gain of voice distinctiveness $G_{VD}$ which is used to estimate how well the anonymized voices of different speakers can be distinguished~\cite{vpc2022}. 

We adopt the B1b and T11 participant system from the Voice Privacy Challenge 2022, in addition to the system proposed by Champion et al.~\cite{pierre_disent} as baselines. 
System T11 is the non-TTS system that achieved the highest privacy level in the 2022 challenge.\footnote{Results available at \url{www.voiceprivacychallenge.org/results-2022}. The overall highest privacy level was in fact achieved by a TTS-based system~\cite{T04} that barely passed the prosody preservation requirement of scoring $\boldsymbol{\rho}^{F_0} > 0.3$. In general, TTS-based systems are known to almost completely erase speaker information, at the cost of a severe loss of intonation and prosody. Therefore, we do not include~\cite{T04} in our comparative analysis.} The work in~\cite{pierre_disent} was the first to propose the use of codebook-based feature quantization.

\section{Results}
\label{sec:results}
Results are shown separately for LibriSpeech and VCTK test sets in~\tabref{tab:obj_results}.
Our system achieves the highest privacy levels: 28.5\% EER for LibriSpeech; 45.5\% EER for VCTK.
The substantially lower EERs of 17.5\% and 28.0\% for the two test sets and the system of Champion et al.\ suggest that our quantization approach is more effective in removing speaker information than that proposed in~\cite{pierre_disent}.
In an effort to further suppress speaker information, Champion et al.\ also experimented with the addition of Gaussian noise to the input F0 curve.  
Improvements to privacy
nonetheless result in a lower pitch correlation $\boldsymbol{\rho}^{F0} \approx 0.55$.\footnote{This result is provided by courtesy of the main author of~\cite{pierre_disent}.} 
For our method, the pitch correlation is in the order of
$\boldsymbol{\rho}^{F0} \approx 0.7$ on average, and compares favorably with that of other systems in the literature~\cite{vpc2022}.
In terms of privacy protection, our model also comfortably outperforms the T11 system by 8\% and 5\% EER for LibriSpeech and VCTK test sets respectively.
The gain in voice distinctiveness for the T11 system are also low.
This is not surprising since the system maps all speakers to similar pseudo-speakers.
In contrast, our system gives values of $G_{VD} \approx -2$, denoting substantially better speaker distinctiveness.

However, utility estimates for our model are lower than that of other systems.
The WER increase from 4.2\% (original data) to 7.5\% for the LibriSpeech subset and from 12.8\% to 18.9\% for the VCTK subset.
Similar issues
have also been reported in the literature.
The authors of~\cite{audiolm} show that the NAC copy-synthesis of LibriSpeech test-clean dataset
causes an increases to the WER of its own ASR system 
from 2.5\% to 6\%, with similar results being reported in~\cite{vall-e}. 
Nevertheless, informal listening tests on our data do not reveal any notable artifacts or degradation to intelligibility.

In an attempt to shed light on the cause for this phenomenon, we
repeated similar experiments using a different ASR architecture, namely that
reported in~\cite{nespoli23_interspeech}, which is retrained according to the same setup described in~\secref{sec:experimental_setup}.
The issue persists.
The WER increases from 2.5\% to 4.6\% for the LibriSpeech subset and from 7.6\% to 15.5\% for the VCTK subset. 
These findings suggest that the degradation to utility is more dependent on the NAC language model than on the ASR system.
As suggested in~\cite{audiolm}, this could be due to the quality of some pseudo-speaker prompts, since the extracted fine acoustic tokens tend also to capture aspects of the (potentially poor) \emph{recording conditions}, the characteristics of which are then transferred to anonymized outputs.  
More thorough experimentation to help us better understand this phenomena is already underway.

\section{Conclusions}
We present a novel approach to speaker anonymization based on a neural audio codec (NAC) language model. Our system performs voice conversion by extracting a set of semantic tokens from an input signal and using them to estimate a set of acoustic tokens belonging to a different speaker, which in turn are used to synthesize an anonymized speech signal with a NAC decoder.
The quantized nature of the semantic and acoustic tokens successfully bottlenecks speaker-related information delivering substantially improved anonymization performance
without compromising prosody or speaker distinctiveness.
While informal listening tests show that anonymized signals are of high quality and intelligibility, automatic transcription with a speech recognition system shows a modest degradation to utility.  Future work should investigate strategies to better protect utility while retaining the benefits to privacy safeguard, such as using high-quality speaker prompts or fine-tuning parts of the system with utility preservation constraints.

\bibliographystyle{IEEEbib}
{\small \balance
\bibliography{refs}}

\end{document}

%% file: symbols.tex
\newcommand{\x}{\mathbf{x}}
\newcommand{\s}{\mathbf{s}}
\newcommand{\ac}{\mathbf{a}}
\newcommand{\prompt}{\mathbf{\Tilde{\ac}}} 
\newcommand{\prediction}{\ac}

\newcommand{\secref}[1]{Section~\ref{#1}}
\newcommand{\tabref}[1]{Table~\ref{#1}}
\newcommand{\figref}[1]{Figure~\ref{#1}}

\newcommand{\michele}[1]{\textcolor{orange}{\textbf{Michele:} #1}}
\newcommand{\attention}[1]{\textcolor{red}{#1}}

\definecolor{light-gray}{gray}{0.925}

%% file: table_res.tex
\begin{table*}[t]
\centering
\small
\begin{tblr}{
  row{2} = {c},
  cell{1}{1} = {r=2}{},
  cell{1}{2} = {c=4}{c},
  cell{1}{6} = {c=4}{c},
  cell{3-12}{2-9} = {c},
  vline{6} = {1-10}{},
  hline{5,6,7,9} = {-}{},
  hline{1,11} = {-}{1.5pt},
  hline{3} = {1}{-}{},
  hline{3} = {2}{-}{},
  cell{9,10}{1-9} = {light-gray}
}
\textbf{System}   & \textbf{\textit{LibriSpeech}} &     &      &      & \textbf{\textit{VCTK}} &      &      &     \\
         & EER (\%) $\uparrow$                     & WER (\%) $\downarrow$ & $G_{VD} \uparrow$ & $\boldsymbol{\rho}^{F_0} \uparrow$ & EER (\%) $\uparrow$         & WER (\%) $\downarrow$ & $G_{VD} \uparrow$ & $\boldsymbol{\rho}^{F_0} \uparrow$ \\
Original & 4.4                           & 4.2 & \SetCell[r=2]{} 0    & \SetCell[r=2]{} 1    & 3.2                    & 12.8 & \SetCell[r=2]{} 0    & \SetCell[r=2]{} 1   \\
Original (eval. pipeline of~\cite{nespoli23_interspeech}) & 1.5 & 2.5 &  &       & 1.1 & 7.6 &  &  \\
B1b~\cite{vpc2022}      & 8.6                           & 4.4 & -5.8 & 0.78 & 9.7                    & 10.7 & -7.1 & 0.81 \\
T11~\cite{t11} & 20.6                          & 3.9 & -19.0 & 0.68 & 39.7                  & 7.9  & -18.4 & 0.73 \\
Champion et al.~\cite{pierre_disent}            & 17.5 & 4.5  & n.a. & 0.67  & 28.0          & 10.0 & n.a. & 0.73 \\
Champion et al. (noise on F0)~\cite{pierre_disent}    & 23.4 & 4.6 & n.a. & 0.52  & 40.8          & 10.3 & n.a. & 0.60 \\
Ours    & 28.5                             & 7.5   & -1.5    & \SetCell[r=2]{} 0.68    & 45.5                      & 18.9    & -2.1    & \SetCell[r=2]{} 0.74    \\
Ours (eval. pipeline of~\cite{nespoli23_interspeech})   & 34.1                             & 4.6   & n.a.    &     & 36.6                      & 15.5   & n.a.    &     \\
\end{tblr}
\caption{Results of the analyzed systems on the Voice Privacy Challenge 2022 test subsets.}
\label{tab:obj_results}
\end{table*}

%% file: Template.bbl
\begin{thebibliography}{10}

\bibitem{introducing_vp}
N.~Tomashenko, Brij Mohan~Lal Srivastava, Xin Wang, Emmanuel Vincent, Andreas Nautsch, Junichi Yamagishi, Nicholas Evans, Jose Patino, Jean-François Bonastre, Paul-Gauthier Noé, and Massimiliano Todisco,
\newblock ``{Introducing the VoicePrivacy Initiative},''
\newblock in {\em Proc. Interspeech 2020}, 2020, pp. 1693--1697.

\bibitem{vpc2022}
Natalia Tomashenko, Xin Wang, Xiaoxiao Miao, Hubert Nourtel, Pierre Champion, Massimiliano Todisco, Emmanuel Vincent, Nicholas Evans, Junichi Yamagishi, and Jean-François Bonastre,
\newblock ``The voiceprivacy 2022 challenge evaluation plan,'' 2022.

\bibitem{fang19}
Fuming Fang, Xin Wang, Junichi Yamagishi, Isao Echizen, Massimiliano Todisco, Nicholas Evans, and Jean-Francois Bonastre,
\newblock ``{Speaker Anonymization Using X-vector and Neural Waveform Models},''
\newblock in {\em Proc. 10th ISCA Workshop on Speech Synthesis (SSW 10)}, 2019, pp. 155--160.

\bibitem{ecapa}
Brecht Desplanques, Jenthe Thienpondt, and Kris Demuynck,
\newblock ``{ECAPA-TDNN: Emphasized Channel Attention, Propagation and Aggregation in TDNN Based Speaker Verification},''
\newblock in {\em Proc. Interspeech 2020}, 2020, pp. 3830--3834.

\bibitem{pierre_disent}
Pierre Champion, Anthony Larcher, and Denis Jouvet,
\newblock ``Are disentangled representations all you need to build speaker anonymization systems?,''
\newblock in {\em Interspeech 2022}. Sept. 2022, pp. 2793--2797, ISCA.

\bibitem{f0_pierre}
Pierre Champion, Denis Jouvet, and Anthony Larcher,
\newblock ``{A Study of F0 Modification for X-Vector Based Speech Pseudonymization Across Gender},'' 2021.

\bibitem{dp}
Ali~Shahin Shamsabadi, Brij Mohan~Lal Srivastava, Aur{\'e}lien Bellet, Nathalie Vauquier, Emmanuel Vincent, Mohamed Maouche, Marc Tommasi, and Nicolas Papernot,
\newblock ``Differentially private speaker anonymization,''
\newblock {\em Proceedings on Privacy Enhancing Technologies}, vol. 1, pp. 98--114, 2023.

\bibitem{audiolm}
Zalán Borsos, Raphaël Marinier, Damien Vincent, Eugene Kharitonov, Olivier Pietquin, Matt Sharifi, Dominik Roblek, Olivier Teboul, David Grangier, Marco Tagliasacchi, and Neil Zeghidour,
\newblock ``{AudioLM}: a {Language} {Modeling} {Approach} to {Audio} {Generation},'' July 2023,
\newblock arXiv:2209.03143 [cs, eess].

\bibitem{vall-e}
Chengyi Wang, Sanyuan Chen, Yu~Wu, Ziqiang Zhang, Long Zhou, Shujie Liu, Zhuo Chen, Yanqing Liu, Huaming Wang, Jinyu Li, Lei He, Sheng Zhao, and Furu Wei,
\newblock ``Neural {Codec} {Language} {Models} are {Zero}-{Shot} {Text} to {Speech} {Synthesizers},'' Jan. 2023,
\newblock arXiv:2301.02111 [cs, eess].

\bibitem{t11}
Jixun Yao, Qing Wang, Li~Zhang, Pengcheng Guo, Yuhao Liang, and Lei Xie,
\newblock ``{NWPU-ASLP System for the VoicePrivacy 2022 Challenge},''
\newblock in {\em Proc. 2nd Symposium on Security and Privacy in Speech Communication}, 2022.

\bibitem{T04}
Sarina Meyer, Pascal Tilli, Florian Lux, Pavel Denisov, Julia Koch, and Ngoc~Thang Vu,
\newblock ``{Cascade of phonetic speech recognition, speaker embeddings gan and multispeaker speech synthesis for the VoicePrivacy 2022 Challenge },''
\newblock in {\em Proc. 2nd Symposium on Security and Privacy in Speech Communication}, 2022.

\bibitem{ohnn}
Xiaoxiao Miao, Xin Wang, Erica Cooper, Junichi Yamagishi, and Natalia Tomashenko,
\newblock ``Language-independent speaker anonymization using orthogonal householder neural network,''
\newblock {\em arXiv preprint arXiv:2305.18823}, 2023.

\bibitem{drift2}
Michele Panariello, Massimiliano Todisco, and Nicholas Evans,
\newblock ``{Vocoder drift compensation by x-vector alignment in speaker anonymisation},''
\newblock in {\em Proc. 3nd Symposium on Security and Privacy in Speech Communication}, 2023.

\bibitem{drift1}
Michele Panariello, Massimiliano Todisco, and Nicholas Evans,
\newblock ``{Vocoder drift in x-vector–based speaker anonymization},''
\newblock in {\em Proc. INTERSPEECH 2023}, 2023, pp. 2863--2867.

\bibitem{soundstream}
Neil Zeghidour, Alejandro Luebs, Ahmed Omran, Jan Skoglund, and Marco Tagliasacchi,
\newblock ``{SoundStream}: {An} {End}-to-{End} {Neural} {Audio} {Codec},''
\newblock {\em IEEE/ACM Transactions on Audio, Speech and Language Processing}, vol. 30, pp. 495--507, Nov. 2021.

\bibitem{encodec}
Alexandre Défossez, Jade Copet, Gabriel Synnaeve, and Yossi Adi,
\newblock ``High {Fidelity} {Neural} {Audio} {Compression},'' Oct. 2022,
\newblock arXiv:2210.13438 [cs, eess, stat].

\bibitem{wang2023lmvc}
Zhichao Wang, Yuanzhe Chen, Lei Xie, Qiao Tian, and Yuping Wang,
\newblock ``Lm-vc: Zero-shot voice conversion via speech generation based on language models,'' 2023.

\bibitem{nespoli23_interspeech}
Francesco Nespoli, Daniel Barreda, Jöerg Bitzer, and Patrick~A. Naylor,
\newblock ``{Two-Stage Voice Anonymization for Enhanced Privacy},''
\newblock in {\em Proc. INTERSPEECH 2023}, 2023, pp. 3854--3858.

\bibitem{hubert}
Wei-Ning Hsu, Benjamin Bolte, Yao-Hung~Hubert Tsai, Kushal Lakhotia, Ruslan Salakhutdinov, and Abdelrahman Mohamed,
\newblock ``Hubert: Self-supervised speech representation learning by masked prediction of hidden units,''
\newblock {\em IEEE/ACM Trans. Audio, Speech and Lang. Proc.}, vol. 29, pp. 3451–3460, oct 2021.

\bibitem{lstm_oh_my_god_am_i_really_citing_this}
Sepp Hochreiter and J\"{u}rgen Schmidhuber,
\newblock ``Long short-term memory,''
\newblock {\em Neural Comput.}, vol. 9, no. 8, pp. 1735–1780, nov 1997.

\bibitem{gpt2}
Alec Radford, Jeffrey Wu, Rewon Child, David Luan, Dario Amodei, Ilya Sutskever, et~al.,
\newblock ``Language models are unsupervised multitask learners,''
\newblock {\em OpenAI blog}, vol. 1, no. 8, pp. 9, 2019.

\bibitem{librispeech}
Vassil Panayotov, Guoguo Chen, Daniel Povey, and Sanjeev Khudanpur,
\newblock ``{Librispeech: An ASR corpus based on public domain audio books},''
\newblock in {\em 2015 IEEE International Conference on Acoustics, Speech and Signal Processing (ICASSP)}, 2015, pp. 5206--5210.

\bibitem{vctk}
Junichi Yamagishi, Christophe Veaux, and Kirsten MacDonald,
\newblock ``{CSTR VCTK Corpus: English Multi-speaker Corpus for CSTR Voice Cloning Toolkit},''
\newblock {\em University of Edinburgh. The Centre for Speech Technology Research}, 2019.

\bibitem{tdnn}
Vijayaditya Peddinti, Daniel Povey, and Sanjeev Khudanpur,
\newblock ``{A time delay neural network architecture for efficient modeling of long temporal contexts},''
\newblock in {\em Proc. Interspeech 2015}, 2015, pp. 3214--3218.

\end{thebibliography}
